\documentclass[]{emulateapj}
\pdfoutput=1
\usepackage{booktabs,threeparttable,graphicx}
\usepackage{CJKutf8}

\begin{document}

\title{Infrared Tip of the Red Giant Branch and Distances to the Maffei/IC~342 Group}

\author{Po-Feng Wu \begin{CJK*}{UTF8}{bsmi}(吳柏鋒)\end{CJK*}\altaffilmark{1}, R. Brent Tully\altaffilmark{1}, Luca Rizzi\altaffilmark{2}, Andrew E. Dolphin\altaffilmark{3}, Bradley A. Jacobs\altaffilmark{1}, Igor D. Karachentsev\altaffilmark{4}}

\altaffiltext{1}{University of Hawaii, Institute for Astronomy, 2680 Woodlawn Drive, HI 96822, USA}
\altaffiltext{2}{W.M. Keck Observatory, 65-1120 Mamalahoa Hwy, Kamuela, HI 96743, USA}
\altaffiltext{3}{Raytheon, 1151 E. Hermans Road, Tucson, AZ 85756, USA}
\altaffiltext{4}{Special Astrophysical Observatory, Russian Academy of Sciences, Nizhnij Arkhyz, Karachai-Cherkessian Republic 369167, Russia}

\begin{abstract}

In this paper, we extend the use of the tip of the red giant branch (TRGB) method to near-infrared wavelengths from previously-used $I$-band, using the \textit{Hubble Space Telescope (HST)} Wide Field Camera 3 (WFC3). Upon calibration of a color dependency of the TRGB magnitude, the IR TRGB yields a random uncertainty of $\sim 5\%$ in relative distance. The IR TRGB methodology has an advantage over the previously-used ACS $F606W$ and $F814W$ filter set for galaxies that suffer from severe extinction. Using the IR TRGB methodology, we obtain distances toward three principal galaxies in the Maffei/IC~342 complex, which are located at low Galactic latitudes. New distance estimates using the TRGB method are 3.45$^{+0.13}_{-0.13}$~Mpc for IC~342, 3.37$^{+0.32}_{-0.23}$~Mpc for Maffei~1 and 3.52$^{+0.32}_{-0.30}$~Mpc for Maffei~2. The uncertainties are dominated by uncertain extinction, especially for Maffei~1 and Maffei~2. Our IR calibration demonstrates the viability of the TRGB methodology for observations with 
the \textit{James Webb Space Telescope (JWST)}.

\end{abstract}

\keywords{galaxies: distances and redshifts --- galaxies: stellar content --- stars: Population II }

\section{Introduction}

Low-mass stars ($\lesssim 2$M$_{\odot}$) begin their main sequence life by burning hydrogen in their cores. As nuclear reactions turn hydrogen into helium, the hydrogen fraction in the core drops and the main energy production takes place in a hydrogen shell. The star gradually becomes brighter, and redder. When it reaches the fully convective Hayashi limit, the color changes only slightly as the star evolves. The nearly vertical track on the color-magnitude diagram (CMD) is the so-called red giant branch (RGB). As a star climbs up the RGB, the burning hydrogen shell dumps helium ashes into the degenerate helium core, increases the core temperature until $\sim 10^8$ K, at which point the triple-$\alpha$ helium burning process can ignite. As soon as the triple-$\alpha$ process is triggered, the star quickly moves away from the RGB. Observationally, this leads to a sharp discontinuity of the luminosity function of the RGB of a galaxy \citep{baa44}. Since the physical condition of triggering the triple-$\alpha$ 
process is well-studied, the luminosity of the star at the tip of RGB (TRGB), therefore the discontinuity, is predictable and can serve as a standard candle for measuring distances of galaxies.  

Stellar evolution theory indicates that the luminosity of the TRGB depends only a little on ages but significantly on the metallicity of the underlying stellar population \citep{ibe83}. Observationally, the age/metallicity dependencies translate into a color dependence of luminosity of the TRGB, which needs to be calibrated empirically. In practice, a TRGB distance measurement requires observations in two bands. One is used to measure the discontinuity of the RGB luminosity function, while the other provides color information that separates RGB stars from bluer main sequence stars, as well as calibrates the age/metallicity variance of the luminosity of the TRGB. 

Currently, the TRGB method is mostly carried out in $I$-band, as the luminosity of TRGB at this wavelength is relatively insensitive to metallicity \citep{sal00,mar08,bel04}, so uncertainties resulting from the empirical color calibration can be minimized. After calibrating the color dependencies of luminosity of the TRGB, \citet{riz07} has demonstrated that, with a single \textit{Hubble Space Telescope} (HST) orbit with the Advance Camera for Surveys (ACS), the distance of a galaxy within $\sim$ 10 Mpc can be obtained. Distances agree well with values derived from Cepheid variables.

In spite of the great accuracy achieved by the optical TRGB methodology, there are still reasons to extend the TRGB method to infrared (IR) wavelengths. Through the optical TRGB method, and other high quality distance measurements, the detailed spatial distribution of unobscured galaxies in the nearby universe is being mapped out \citep{tul13}. The three-dimensional spatial information, together with the radial velocities of these galaxies, has provided precious information for dynamical studies probing the distribution of dark matter and dark energy in the nearby universe. However, while galaxies at low Galactic latitudes are dynamically equally important, their distribution remains under-explored due to severe obscuration. IR wavelengths offer a solution because the TRGB in IR is brighter and Galactic extinction is reduced. With an accurate calibration of TRGB magnitudes in the IR, distances of obscured galaxies can be obtained, providing a more complete three-dimensional map of the nearby universe. In 
addition, while HST is the current work horse for the TRGB methodology, its successor the James Webb Space Telescope (JWST) functions only in the IR. 
It will be necessary to migrate procedures from optical to IR wavelengths. Once JWST is operational, the capabilities of the TRGB method will be enhanced dramatically, allowing accurate distances to be measured to galaxies within the Virgo Cluster and beyond with short exposures. Thousands of galaxies will be within reach. 

Already with Wide Field Camera 3 (WFC3) installed on HST, the facility exists to work in the infrared domain. An important demonstration of its possibilities is discussed by \citet{dal12b}. With WFC3 $F110W$ and $F160W$ filters, which approximately correspond to $J$ and $H$ bands, these authors observed 23 nearby galaxies ($2.0 \mbox{ Mpc} \lesssim D \lesssim 4.5 \mbox{ Mpc}$), which had already been studied in detail at HST optical bands, mostly with ACS. That paper includes a study of the dependencies of the luminosity of the TRGB, with cognizance of the potential utility as a distance indicator. 
However, \citet{dal12b} focus on the TRGB in the $F160W$ band where luminosities are the greatest at wavelengths accessible to WFC3 but metallicity effects are substantial. Our interest in the current study is to calibrate the TRGB methodology in the $F110W$ band. It will be demonstrated that changes in the luminosity of the TRGB with color are cut in half at $F110W$ compared with $F160W$. A color calibration is required for working at $F110W$ but uncertainties are reduced.

In this paper, we first present the calibration of the color dependencies of TRGB magnitudes in both $F110W$ and $F160W$. We then apply our calibration to obtain distances of galaxies in the Maffei/IC~342 complex, an entity that is important because it produces the greatest tidal influence on the Local Group \citep{dun93}. We describe the data used in this paper and the data reduction in Section~2. Calibration of the color dependency of TRGB magnitudes is presented in Section~3. We compare distances derived from IR TRGB and optical TRGB methods is Section~4. Processes of measuring distance to members of Maffei/IC~342 groups are discussed in Section~5. Section~6 gives the summary. All magnitudes in this paper are Vega magnitudes.

\section{Data} 
\label{sec:data}

\subsection{Calibration Sample}
Data for calibrating the color dependencies of IR TRGB magnitudes are retrieved from two sources. The first data set is from the HST Program 11719 (P.I.: J. Dalcanton), in which 26 fields in 23 nearby galaxies were observed in both $F110W$ and $F160W$ filters with $HST$ WFC3. This set of galaxies is ideal for our purpose because of the way they were selected. First, they have sufficient numbers of stars so that their RGB are well populated within a single WFC3 field of view. Second, the stellar populations are not over-crowded, so photometry of good quality can be carried out. Third, the sample spans a wide range of TRGB colors, hence provides a wide metallicity-age baseline for our calibration. Fourth, all these galaxies have distances derived from $F814W$ TRGB magnitudes, which serve as good calibrators for IR TRGB magnitudes. For detailed positioning of the observations, we refer the reader to \citet{dal12b}.

The 26 fields in these 23 galaxies cover a wide metallicity range of $-2.0 \lesssim [Fe/H] \lesssim -0.4$ \citep{dal12b,mel12}, that provides an extended metallicity-age baseline for calibration purposes. To further extend the baseline at the high metallicity end, we add observations of M31 from the Panchromatic Hubble Andromeda Treasury \citep[PHAT,][]{dal12a}. M31, especially its bulge, will provide information about high metallicity stars \citep{ols06}. PHAT images about one-third of the star-forming disk of M31 in six filters, including $F110W$ and $F160W$ with WFC3, as well as $F475W$ and $F814W$ with ACS. For detailed design of the survey, we refer readers to \citet{dal12a}. Here we briefly describe survey arrangements relevant to our study. 

The PHAT survey area is divided into 23 sub-areas, called ``bricks``. We select bricks along the major axis of M31, roughly equally spread from the center of the bulge to outer disk in order to sample the full metallicity range provided by M31. We avoid bricks located at M31's star-forming arms, as a prominent stellar population with age $\lesssim 2$ Gyrs may interfere with TRGB determinations \citep{bar04,sal05}. As a result, bricks B01, B05, B09, B13, B19 and B23 are selected. Each brick is composed of 18 pointings, each called a ``field``.  Because a single pointing already contains enough stars for a TRGB determination, we arbitrarily chose only Field 1 in each brick for our measurements. Thereby, we add six measurements to our calibration sample, intended to augment the representation at the high metallicity end.

\subsection{Maffei/IC~342 Complex}

\begin{figure}
 \includegraphics[width=\columnwidth,]{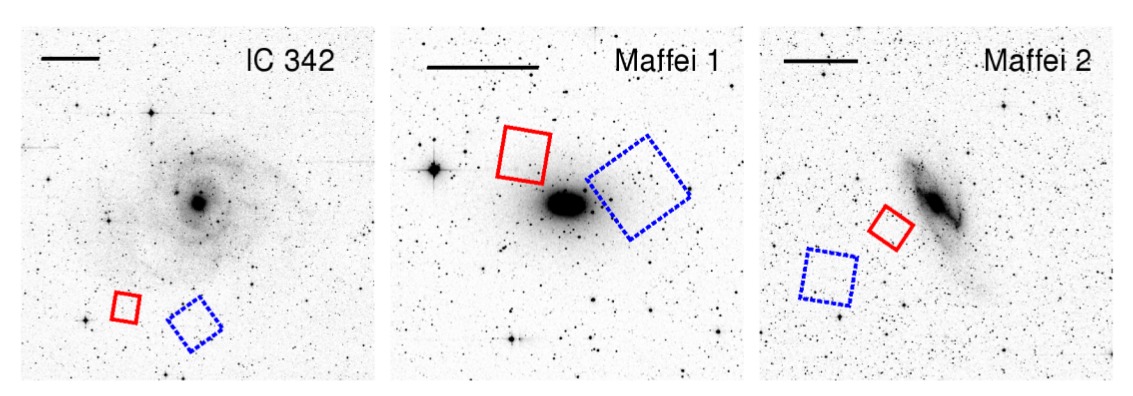}
\caption{Positions of the WFC3/IR (red solid box) and ACS (blue dashed box) fields of view, overlaid on 2MASS $K_s$-band images of IC~342, Maffei~1 and Maffei~2. Black lines in the upper left corners represent an angular size of 5\arcmin. Assuming a distance of 3.5~Mpc to the Maffei-IC342 group, 5\arcmin corresponds to $\sim 5$~kpc.}
\label{fig:footprint}
\end{figure}

Observations were carried out during $HST$ Cycle 20 as Program 12877 (PI: I. D. Karachentsev). We obtained observations of three principal and highly obscured galaxies in this nearest group (D $\simeq 3$Mpc): IC~342 (Scd), Maffei~1 (E), and Maffei~2 (Sbc). Each galaxy was observed with two orbits, with WFC3 through the $F110W$ and $F160W$ filters as primary and ACS through $F606W$ and $F814W$ filters as parallel. The total exposure time is 2411.7s with $F110W$ and $F160W$, and 1946.0s with $F606W$ and $F814W$ for each galaxy.
Figure~\ref{fig:footprint} shows the locations where we pointed the cameras toward the three galaxies, overlaid on the Two Micron All Sky Survey \citep[2MASS,][]{skr06} $K_s$-band images. We pointed the WFC3/IR camera a few arc-minutes away from the galaxy center to avoid potential crowding issues that would affect precise photometry. Positions of the ACS camera were chosen semi-arbitrarily, only to avoid crowded \ion{H}{2}-regions in the galaxy as well as bright foreground stars. 

\subsection{Photometry}
\label{sec:photometry}

Photometry is performed using the package DOLPHOT 2.0\footnote{http://americano.dolphinsim.com/dolphot/} \citep{dol00} with either the WFC3 or ACS module for corresponding observations. We adopt processing parameters suggested by the DOLPHOT manual except for M31. We change relevant processing parameters for M31 because at the surface density of M31, crowding may be a problem in photometry, especially in the bulge. First, we use a smaller photometry aperture radius \texttt{RAper} = 2 instead of the default value of 4 for ACS and 3 for WFC3. The smaller photometry aperture radius reduces contamination from neighboring unresolved stars therefore resulting in more precise photometry. Second, we set \texttt{FitSky} = 2, as suggested to be optimal for extremely crowded fields by the DOLPHOT manual, instead the default value of 1. 
 
The DOLPHOT output assigns several quality parameters for each measurement. Following suggestions in the DOLPHOT manual, we accept measurements to be reliable if $-0.3 < $ \texttt{Sharpness} $<0.3$, object type = 1 or 2 (stars), photometric quality flag = 0 and signal-to-noise ratio (S/N) $\ge$ 5 in both $F110W$/$F160W$, $F606W$/$F814W$ or $F475W$/$F814W$ filters.

We assess the completeness, uncertainty and bias of our photometry as a function of magnitudes using artificial star tests. We place a series of 100,000 artificial stars into the image, with magnitudes and colors of stars that cover the range of interest. We insert one artificial star at a time, then rerun the photometry using the identical photometry procedures to determine if the star is found. If the star is detected, then it also returns all photometric parameters as in real photometry. The output then goes through the identical quality cuts as applied to real measurements. The result is then binned by the magnitudes of input artificial stars. The photometric completeness is determined as the ratio between stars recovered and stars inserted at a given magnitude. In each magnitude bin, the distribution of recovered magnitudes is fitted by a Gaussian. 
The center and dispersion of the fitted Gaussian is determined as the photometric bias and uncertainty at the given magnitude. This information is incorporated into the TRGB determination, which will be described in the next section.

\section{Calibration of TRGB Magnitudes in the NIR}
\label{sec:cali}

We derive TRGB apparent magnitudes by fitting a pre-defined model luminosity function to the data, following the approach presented by \citet{mak06}.

The model luminosity function is parameterized by two distinct power-laws, where the discontinuity of the luminosity function represents the TRGB magnitude:

\begin{equation}
 \psi = \left\{ \begin{array}{lr}
                 10^{a(m-m_{TRGB})+b}, & m - m_{TRGB} \ge 0, \\
                 10^{c(m-m_{TRGB})},   & m - m_{TRGB} < 0.
                \end{array} \right.
\end{equation}

This pre-defined luminosity function is then convolved with the completeness, uncertainty and bias:

\begin{equation}
 \varphi(m) = \int \psi(m') \rho(m') e(m|m')dm',
\end{equation}

where $\rho(m)$ is the completeness as a function of magnitude and $e(m|m')$ is the error distribution function, both acquired from artificial star tests. The error distribution function contains photometric uncertainty and bias:

\begin{equation}
e(m|m') = \frac{1}{\sqrt{2\pi}\sigma(m')} \exp \left\{ - \frac{[m - \bar{m}(m')]^2}{2\sigma^2(m')} \right\},
\end{equation}

where $\sigma(m)$ is the uncertainty and $(\bar m)$ is the bias. The luminosity function after convolving with completeness, bias and uncertainty, $\varphi$, then becomes the luminosity function fitted to the data.

\begin{figure}
 \includegraphics[width=\columnwidth,]{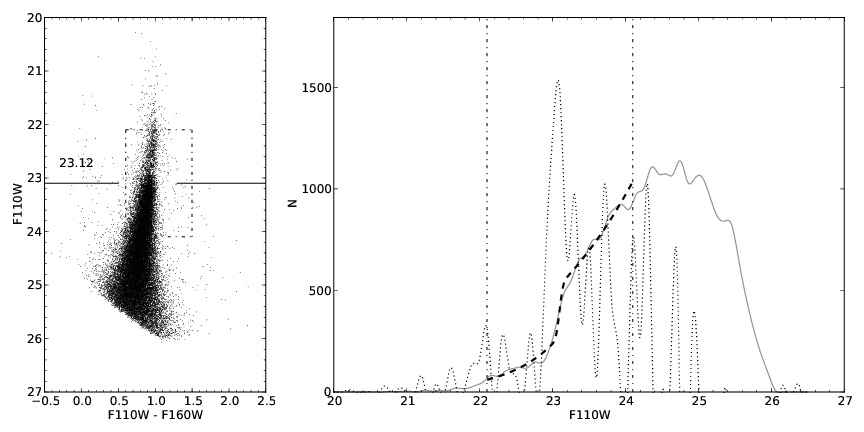}
 \caption{The CMD and $F110W$ luminosity function of DDO82 as an example of the TRGB fitting procedure. Both are before extinction correction. {\it Left:} CMD of DDO82. The TRGB is indicated by the horizontal line. Only stars in the box are included for TRGB fitting. {\it Right:} $F110W$ luminosity function of DDO82. Only stars located within the range of the box on the CMD are included. The peak of the dotted line, the first derivative of luminosity function, provides a good initial guess of the TRGB magnitude. The gray solid line is the smoothed luminosity function. The luminosity function was fitted by a step power law convolved with photometric errors and incompleteness. Two vertical dash-dotted lines indicate $\pm$1 mag of the initial guess of TRGB magnitude, where the fitting was carried out. The best-fit result is plotted as the thick dashed line.}
 \label{fig:CMD}
\end{figure}

We bin the data every 0.05 mag to construct the observed luminosity function. We use the peak of the first derivative of the observed luminosity function as our first guess of the TRGB magnitude, then fit $\varphi$ to the observed luminosity function over a range of $\pm$1 mag of the first guess value of the TRGB (see Figure~\ref{fig:CMD} for an example). The fitting procedure is based on a non-linear least squares method, as it requires less computational resources, and uses the Levenberg-Marquardt algorithm to find the best-fit parameters. The square-root of the variance of fitted TRGB magnitude is taken as the uncertainty of measurement of the TRGB magnitude. 
The color of the TRGB is derived from the median color of stars within 0.05 mag fainter than the TRGB. This magnitude range provides a measurement of color close to the TRGB, and usually contains enough stars ($> 20$) for a robust color measurement. A few galaxies (UGCA292, ESO540-030, HS117, KDG73) have less then 20 stars in the first 0.05 mag bin below the TRGB which reflects in their more uncertain color measurements (see Table~\ref{tbl:galaxydata}). We access the uncertainty of TRGB color from 1000 bootstrap resampling trials. We report the 16th and 84th percentile of the 1000 trials as the range of the color uncertainty. The procedure is run for both $F110W$ and $F160W$ luminosity functions to derive TRGB magnitudes in both bands.

\begin{figure*}
 \includegraphics[width=\textwidth,]{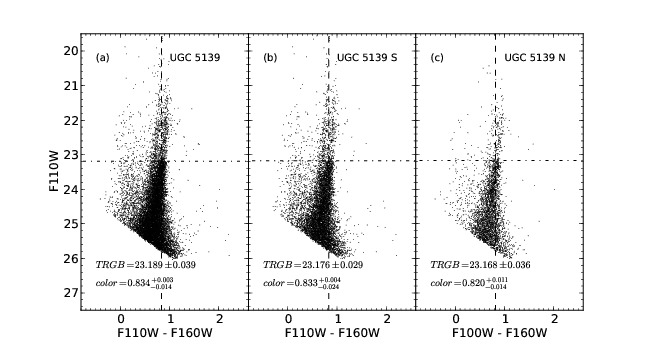}
 \caption{IR CMDs and TRGB measurements of UGC~5139. Panel (a) shows the CMD of UGC~5139. A prominent RHeB population is located at $\lesssim 0.2$ mag blueward of RGB. The field is split into two sub-regions, the southern half (b), where the RHeB stars are located and the northern half (c), which is relatively free from RHeB stars. The same TRGB measurement procedure is carried out on all three fields. The horizontal dash-dotted line represents the TRGB magnitudes while the vertical dashed line indicates the TRGB colors. Both the TRGB magnitudes and colors in the two sub-regions agree with each other, also with values derived from the whole UGC~5139 within measurement uncertainties.}
 \label{fig:ugc5139}
\end{figure*}

For the calibration sample, we restrict our analysis to stars with colors $0.6 < F110W-F160W < 1.5$ to minimize the effect of the blue main-sequence star population. Another stellar population that would potentially affect TRGB detection is the red core helium burning (RHeB) stars located blueward of the RGB, for which a clean separation from the RGB stars is difficult by a simple color cut. Figure~\ref{fig:ugc5139}a shows the $F110W$ v.s. $F110W - F160W$ CMD of UGC~5139, in which the prominent RHeB population is located $\lesssim 0.2$ mag blueward of the RGB population at the magnitude of the TRGB. 
In Figure~\ref{fig:ugc5139}b and c, we apply spatial selections, only including stars in the southern and northern half of the WFC3 coverage, respectively. The RHeB population is mainly located in the southern half, while the northern half is relatively free from RHeB stars. We carry out TRGB measurements on all the data sets. 
Measurements in the two sub-regions and the whole field agree within measurement uncertainties. The RHeB population does not have a noticeable effect on the TRGB determination here. \citet{mak06}, who determined the $F814W$ TRGB using the same form of model luminosity function, found the similar result that the TRGB measurement is nearly unaffected by intermediate-age and/or young stars when the TRGB is much brighter than the photometric limit. 
However, we note that when the TRGB is close to the photometric limit, the existence of a significant young and/or intermediate-age stellar population such as RHeB stars and AGB stars would interfere with the TRGB determination. In difficult cases, spatially filtering out young/intermediate-age stellar population, as shown with UGC~5139, can help isolate the TRGB \citep{mak06}. For our calibration sample, the TRGB is brighter than the photometric limit by $\gtrsim 2$ mag, therefore we consider the effects from young stellar population are negligible and do not apply spatial filtering.

Table~\ref{tbl:galaxydata} summarizes TRGB measurements, along with $E(B-V)$, distance moduli derived from $F814W$ TRGB, and absolute TRGB magnitudes in $F110W$ and $F160W$ with the $F814W$ TRGB distance modulus applied. The magnitudes and colors in Table~\ref{tbl:galaxydata} are not corrected for dust extinction. CMDs of the calibration sample can be found in the Extragalactic Distance Database\footnote{http://edd.ifa.hawaii.edu/} \citep[EDD;][]{tul09}. From the EDD webpage, selecting the ``CMDs/TRGB`` index page in the block ``Stellar Distances`` will show a table containing galaxies with TRGB distance measurements and relevant information. CMD of each galaxy is accessible by clicking its name in the column ``Name/CMD``. A detailed description of the ``CMDs/TRGB`` index page is presented in \citet{jac09}.

\tiny
\begin{table*}
\begin{threeparttable}
\label{tbl:galaxydata}
 \caption{Measurements of Calibration Sample}
\tabcolsep=0.11cm
\begin{tabular}[t]{llrcccccccc}
 \hline
 \hline
PGC & Target & T\tnote{a} & $m_{TRGB}^{F110W}$\tnote{b} & Color\tnote{b} & $m_{TRGB}^{F160W}$\tnote{b} & Color\tnote{b} & $E(B-V)$\tnote{c} & $m-M$\tnote{d} & $M_{TRGB}^{F110W}$ & $M_{TRGB}^{F160W}$ \\
Number & & & & & & & & & \\
\hline
2881 & ESO540-030 & 9 & 22.977$\pm$0.023 & 0.848$^{+0.010}_{-0.011}$ & 22.115$\pm$0.032 & 0.845$^{+0.001}_{-0.030}$ & 0.020 & 27.76$^{+0.04}_{-0.04}$ & -4.801$^{+0.061}_{-0.046}$ & -5.655$^{+0.051}_{-0.051}$ \\
3238 & NGC300 & 7 & 21.626$\pm$0.039 & 0.947$^{+0.008}_{-0.001}$ & 20.602$\pm$0.044 & 0.962$^{+0.002}_{-0.003}$ & 0.011 & 26.60$^{+0.06}_{-0.05}$ & -4.984$^{+0.098}_{-0.063}$ & -6.004$^{+0.074}_{-0.067}$ \\
4126 & NGC404 & -3 & 22.527$\pm$0.084 & 0.968$^{+0.001}_{-0.014}$ & 21.522$\pm$0.020 & 0.977$^{+0.001}_{-0.005}$ & 0.051 & 27.37$^{+0.02}_{-0.02}$ & -4.888$^{+0.168}_{-0.086}$ & -5.874$^{+0.028}_{-0.028}$ \\
21396 & NGC2403-DEEP & 6 & 22.644$\pm$0.075 & 0.931$^{+0.002}_{-0.006}$ & 21.713$\pm$0.045 & 0.944$^{+0.012}_{-0.002}$ & 0.035 & 27.52$^{+0.05}_{-0.05}$ & -4.907$^{+0.157}_{-0.090}$ & -5.825$^{+0.067}_{-0.067}$ \\
21396 & NGC2403-HALO6 & 6 & 22.612$\pm$0.052 & 0.930$^{+0.004}_{-0.023}$ & 21.655$\pm$0.073 & 0.890$^{+0.025}_{-0.011}$ & 0.035 & 27.52$^{+0.05}_{-0.05}$ & -4.939$^{+0.115}_{-0.072}$ & -5.883$^{+0.088}_{-0.088}$ \\
21396 & NGC2403-SN & 6 & 22.612$\pm$0.018 & 0.931$^{+0.004}_{-0.003}$ & 21.590$\pm$0.017 & 0.960$^{+0.002}_{-0.002}$ & 0.035 & 27.52$^{+0.05}_{-0.05}$ & -4.939$^{+0.062}_{-0.053}$ & -5.948$^{+0.053}_{-0.053}$ \\
23324 & UGC4305 & 10 & 22.898$\pm$0.035 & 0.843$^{+0.004}_{-0.008}$ & 21.985$\pm$0.017 & 0.855$^{+0.008}_{-0.019}$ & 0.028 & 27.70$^{+0.02}_{-0.02}$ & -4.827$^{+0.073}_{-0.040}$ & -5.729$^{+0.026}_{-0.026}$ \\
23324 & UGC4305-2 & 10 & 22.805$\pm$0.037 & 0.820$^{+0.008}_{-0.017}$ & 21.966$\pm$0.025 & 0.847$^{+0.009}_{-0.010}$ & 0.028 & 27.70$^{+0.02}_{-0.02}$ & -4.920$^{+0.077}_{-0.042}$ & -5.748$^{+0.031}_{-0.032}$ \\
24050 & UGC4459 & 10 & 23.010$\pm$0.036 & 0.820$^{+0.006}_{-0.005}$ & 22.164$\pm$0.023 & 0.831$^{+0.003}_{-0.001}$ & 0.032 & 27.83$^{+0.03}_{-0.02}$ & -4.848$^{+0.079}_{-0.041}$ & -5.682$^{+0.038}_{-0.030}$ \\
27605 & UGC5139 & 10 & 23.189$\pm$0.039 & 0.834$^{+0.003}_{-0.014}$ & 22.367$\pm$0.050 & 0.854$^{+0.005}_{-0.014}$ & 0.043 & 28.02$^{+0.02}_{-0.03}$ & -4.869$^{+0.081}_{-0.049}$ & -5.675$^{+0.054}_{-0.058}$ \\
28120 & NGC2976 & 5 & 22.912$\pm$0.030 & 0.965$^{+0.003}_{-0.003}$ & 21.890$\pm$0.042 & 0.977$^{+0.007}_{-0.008}$ & 0.062 & 27.82$^{+0.07}_{-0.03}$ & -4.963$^{+0.092}_{-0.042}$ & -5.962$^{+0.082}_{-0.052}$ \\
28630 & M81 & 2 & 22.889$\pm$0.041 & 1.018$^{+0.013}_{-0.018}$ & 21.798$\pm$0.071 & 1.028$^{+0.033}_{-0.023}$ & 0.071 & 27.84$^{+0.09}_{-0.09}$ & -5.014$^{+0.121}_{-0.099}$ & -6.078$^{+0.115}_{-0.115}$ \\
29146 & NGC3077 & 7 & 23.080$\pm$0.041 & 0.994$^{+0.001}_{-0.015}$ & 22.024$\pm$0.060 & 1.009$^{+0.009}_{-0.005}$ & 0.059 & 27.93$^{+0.08}_{-0.06}$ & -4.902$^{+0.115}_{-0.073}$ & -5.936$^{+0.100}_{-0.085}$ \\
29257 & DDO71 & 10 &  23.056$\pm$0.026 & 0.883$^{+0.005}_{-0.005}$ & 22.145$\pm$0.043 & 0.898$^{+0.002}_{-0.001}$ & 0.084 & 27.81$^{+0.01}_{-0.02}$ & -4.828$^{+0.053}_{-0.033}$ & -5.708$^{+0.044}_{-0.047}$ \\
30664 & DDO78 & 10 & 22.961$\pm$0.055 & 0.899$^{+0.001}_{-0.015}$ & 22.064$\pm$0.029 & 0.907$^{+0.004}_{-0.015}$ & 0.020 & 27.71$^{+0.04}_{-0.03}$ & -4.767$^{+0.117}_{-0.063}$ & -5.656$^{+0.049}_{-0.042}$ \\
30819 & IC2574 & 9 & 23.178$\pm$0.026 & 0.847$^{+0.002}_{-0.004}$ & 22.270$\pm$0.034 & 0.866$^{+0.003}_{-0.004}$ & 0.032 & 27.97$^{+0.03}_{-0.02}$ & -4.820$^{+0.060}_{-0.033}$ & -5.716$^{+0.045}_{-0.039}$ \\
30997 & DDO82 & 9 & 23.100$\pm$0.015 & 0.894$^{+0.001}_{-0.007}$ & 22.165$\pm$0.027 & 0.920$^{+0.003}_{-0.004}$ & 0.038 & 27.97$^{+0.01}_{-0.02}$ & -4.903$^{+0.032}_{-0.025}$ & -5.824$^{+0.029}_{-0.034}$ \\
32667 & KDG73 & 10 & 23.265$\pm$0.031 & 0.786$^{+0.002}_{-0.011}$ & 22.498$\pm$0.063 & 0.783$^{+0.014}_{-0.046}$ & 0.016 & 27.96$^{+0.17}_{-0.12}$ & -4.709$^{+0.181}_{-0.124}$ & -5.470$^{+0.181}_{-0.136}$ \\
35878 & NGC3741 & 10 & 22.781$\pm$0.045 & 0.753$^{+0.020}_{-0.002}$ & 21.960$\pm$0.033 & 0.814$^{+0.004}_{-0.008}$ & 0.023 & 27.54$^{+0.11}_{-0.12}$ & -4.779$^{+0.141}_{-0.128}$ & -5.592$^{+0.115}_{-0.124}$ \\
38881 & NGC4163 & 10 & 22.500$\pm$0.043 & 0.840$^{+0.008}_{-0.001}$ & 21.631$\pm$0.026 & 0.867$^{+0.003}_{-0.002}$ & 0.017 & 27.38$^{+0.03}_{-0.02}$ & -4.895$^{+0.091}_{-0.047}$ & -5.758$^{+0.040}_{-0.033}$ \\
42275 & UGCA292 & 10 & 23.228$\pm$0.049 & 0.729$^{+0.007}_{-0.027}$ & 22.659$\pm$0.179 & 0.725$^{+0.036}_{-0.022}$ & 0.014 & 27.93$^{+0.31}_{-0.05}$ & -4.714$^{+0.325}_{-0.070}$ & -5.278$^{+0.358}_{-0.186}$ \\
47495 & UGC8508 & 10 & 22.267$\pm$0.032 & 0.792$^{+0.011}_{-0.016}$ & 21.433$\pm$0.013 & 0.829$^{+0.007}_{-0.008}$ & 0.013 & 27.13$^{+0.08}_{-0.08}$ & -4.874$^{+0.103}_{-0.086}$ & -5.704$^{+0.081}_{-0.081}$ \\
73049 & NGC7793 & 7 & 22.936$\pm$0.057 & 0.920$^{+0.012}_{-0.011}$ & 22.075$\pm$0.076 & 0.931$^{+0.000}_{-0.006}$ & 0.018 & 27.80$^{+0.08}_{-0.08}$ & -4.880$^{+0.139}_{-0.098}$ & -5.734$^{+0.110}_{-0.110}$ \\
95597 & KKH37 & 10 & 22.854$\pm$0.028 & 0.849$^{+0.004}_{-0.002}$ & 22.005$\pm$0.040 & 0.869$^{+0.002}_{-0.009}$ & 0.068 & 27.68$^{+0.04}_{-0.04}$ & -4.886$^{+0.070}_{-0.049}$ & -5.710$^{+0.056}_{-0.057}$ \\
3097727 & Sc-dE1 & -3 & 23.430$\pm$0.056 & 0.780$^{+0.009}_{-0.012}$ & 22.604$\pm$0.074 & 0.787$^{+0.018}_{-0.017}$ & 0.013 & 28.16$^{+0.04}_{-0.03}$ & -4.741$^{+0.119}_{-0.064}$ & -5.563$^{+0.084}_{-0.080}$ \\
4689216 & HS117 & 10 & 23.245$\pm$0.035 & 0.849$^{+0.014}_{-0.003}$ & 22.355$\pm$0.067 & 0.872$^{+0.015}_{-0.019}$ & 0.102 & 27.99$^{+0.02}_{-0.06}$ & -4.835$^{+0.074}_{-0.069}$ & -5.687$^{+0.070}_{-0.090}$ \\
2557 & M31-B01 & 3 & 19.297$\pm$0.019 & 1.090$^{+0.010}_{-0.002}$ & 18.195$\pm$0.016 & 1.108$^{+0.002}_{-0.003}$ & 0.186 & 24.40$^{+0.08}_{-0.08}$ & -5.103$^{+0.082}_{-0.082}$ & -6.205$^{+0.082}_{-0.082}$ \\
2557 & M31-B05 & 3 & 19.299$\pm$0.027 & 1.082$^{+0.006}_{-0.005}$ & 18.193$\pm$0.011 & 1.121$^{+0.002}_{-0.005}$ & 0.206 & 24.40$^{+0.08}_{-0.08}$ & -5.101$^{+0.084}_{-0.084}$ & -6.207$^{+0.081}_{-0.081}$ \\
2557 & M31-B09 & 3 & 19.367$\pm$0.035 & 1.039$^{+0.004}_{-0.003}$ & 18.280$\pm$0.034 & 1.054$^{+0.011}_{-0.012}$ & 0.208 & 24.40$^{+0.08}_{-0.08}$ & -5.033$^{+0.087}_{-0.087}$ & -6.120$^{+0.087}_{-0.087}$ \\
2557 & M31-B13 & 3 & 19.382$\pm$0.048 & 1.040$^{+0.003}_{-0.003}$ & 18.340$\pm$0.043 & 1.050$^{+0.006}_{-0.008}$ & 0.226 & 24.40$^{+0.08}_{-0.08}$ & -5.018$^{+0.093}_{-0.093}$ & -6.060$^{+0.091}_{-0.091}$ \\
2557 & M31-B19 & 3 & 19.486$\pm$0.064 & 1.022$^{+0.004}_{-0.005}$ & 18.446$\pm$0.069 & 1.030$^{+0.007}_{-0.006}$ & 0.198 & 24.40$^{+0.08}_{-0.08}$ & -4.914$^{+0.102}_{-0.102}$ & -5.954$^{+0.106}_{-0.106}$ \\
2557 & M31-B23 & 3 & 19.515$\pm$0.035 & 1.030$^{+0.009}_{-0.009}$ & 18.498$\pm$0.064 & 1.038$^{+0.014}_{-0.005}$ & 0.204 & 24.36$^{+0.08}_{-0.08}$ & -4.885$^{+0.087}_{-0.087}$ & -5.902$^{+0.102}_{-0.102}$ \\
\hline

\end{tabular}

\begin{tablenotes}
 \item[a] Morphological type T from HyperLEDA. http://leda.univ-lyon1.fr .
 \item[b] Before correcting for extinction.
 \item[c] All E(B-V) except for M31 are from \citet{sch11}. For M31, we take values from \citet{mon09}.
 \item[d] Distance moduli are from $F814W$ TRGB distance in EDD, except M31. For M31, we derive the distance modulus using $F814W$ TRGB magnitude and color in EDD, but with extinction map of M31 from \citet{mon09}. See \S~3 for details about EDD and the extinction of M31.
\end{tablenotes}
\end{threeparttable}
\end{table*}

\normalsize

\begin{figure}
 \includegraphics[width=\columnwidth,]{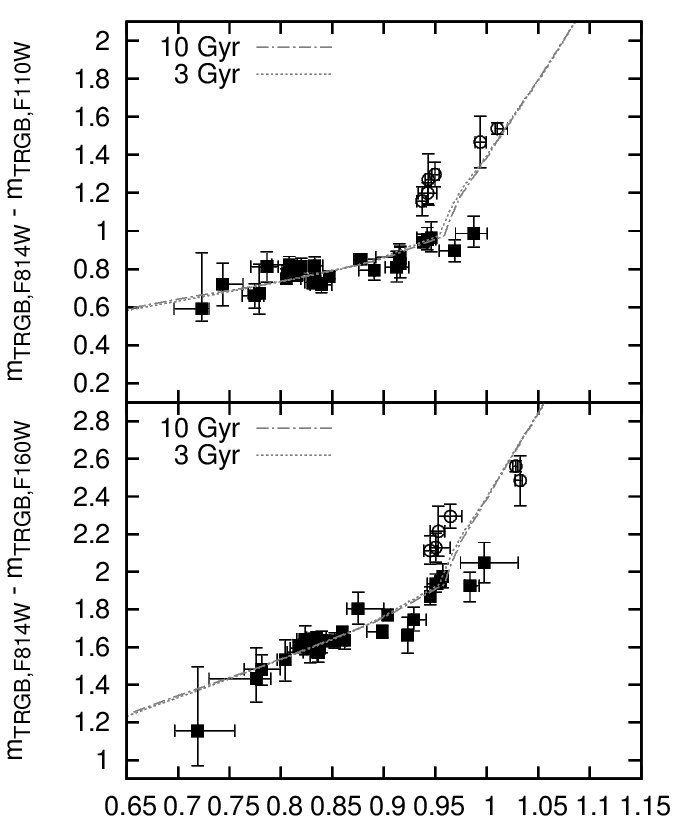}
\caption{Magnitude difference between optical and IR TRGB magnitudes vs. $F110W - F160W$ color. Filled squares represent galaxies in $HST$ Program 11719, while open circles represent M31 from the PHAT survey. Errors in magnitudes are quadratic sums of errors from TRGB measurements in two bands. All data are corrected for extinction. Dash-dotted and dotted lines are from the isochrone models of \citet{bre12} for ages of 10~Gyr and 3~Gyr with metallicity $0.0001 < Z \lesssim 0.02$. The blue end corresponds to the lowest metallicity and the color increases when metallicity increases.}
\label{fig:814110}
\end{figure}

Figure~\ref{fig:814110} shows the magnitude difference between $F814W$ and IR TRGB magnitudes as a function of $F110W - F160W$ color, after correcting for Galactic extinction. The isochrone models of \citet{bre12} for ages of 3~Gyr and 10~Gyr with various metallicities from $Z = 10^{-4}$ up to $Z \simeq 0.02$ are also shown in the plot. For galaxies except M31, the $F814W$ TRGB magnitudes are directly obtained from the EDD. The column ``T814`` of the ``CMDs/TRGB`` table lists $F814W$ TRGB magnitudes before correcting for Galactic extinction, and columns ``T8\_lo`` and ``T8\_hi`` are $1 \sigma$ error estimates. Galactic extinction is based on \citet{sch11}, a recalibration of the \citet{sch98} dust map, acquired through the NASA/IPAC Extragalactic Database (NED). Here we assume that internal extinction is negligible because (1) most of these galaxies are irregular morphological types and (2) for other earlier morphological type galaxies, the WFC3 usually covers their halos region, which is presumably mostly 
dust free \citep[see][for footprints of WFC3 coverage.]{dal12b}. This assumption may not be good for a few fields (NGC~2403-SN, NGC~300, NGC~404), but we do not make corrections for this small portion of the sample. 
For M31, the $F814W$ TRGB magnitudes of each field are derived using the same procedure for IR TRGB, with ACS $F814W$ and $F475$ observations made on the same locations as the PHAT survey (Table~\ref{tab:M31}). We use the extinction map of M31 derived by \citet{mon09} instead of from \citet{sch11} since the extinction values around M31 are not well calibrated in \citet{sch98}. Throughout this paper, we convert extinction among different filters using conversion factors provided by \citet{sch11}, assuming $R_V=3.1$.

\begin{table}[h]
\begin{threeparttable}
\caption{M31 $F814W$ TRGB Measurements}
\begin{tabular}[t]{lccc}
\hline
\hline
Brick & HST Program & $m_{TRGB}^{F814W}$\tnote{a} & $F475W-F814W$\tnote{a} \\
\hline
B01 & 12058 & 20.975$\pm$0.024 & 3.890$^{+0.012}_{-0.012}$ \\
B05 & 12074 & 20.921$\pm$0.133 & 3.810$^{+0.087}_{-0.048}$ \\
B09 & 12057 & 20.820$\pm$0.054 & 3.770$^{+0.098}_{-0.092}$ \\
B13 & 12114 & 20.822$\pm$0.126 & 3.506$^{+0.054}_{-0.152}$ \\
B19 & 12110 & 20.792$\pm$0.040 & 3.391$^{+0.063}_{-0.071}$ \\
B23 & 12070 & 20.867$\pm$0.042 & 3.571$^{+0.077}_{-0.144}$ \\
\hline
\end{tabular}
\begin{tablenotes}
 \item[a] Not corrected for extinction. 
\end{tablenotes}
\label{tab:M31}
\end{threeparttable}
\end{table}

Figure~\ref{fig:814110} shows that $F814W - \mbox{IR}$ colors rise near linearly with $F110W-F160W$ color until taking a sharp upturn at the red end. Theoretical models predict a similar behavior. High metallicity objects have redder TRGB colors, with an upturn that starts at $F110W-F160W \simeq 0.95$. The upturn is a result from noticeable line-blanketing effects at $\mbox{[Fe/H]} \gtrsim -0.7$ ($Z \gtrsim 10^{-3}$), which suppresses the luminosity in $I$-band \citep{bar04,mag08}.

We then transform from apparent to absolute magnitudes using the $F814W$ TRGB distances as sources of input distance moduli (see Table~\ref{tbl:galaxydata}). For each galaxy except M31, we adopt the distance modulus reported in the EDD, which is the column ``DM\_tip`` on the ``CMDs/TRGB`` page. The $F814W$ TRGB distances in EDD were derived using the calibration of \citet{riz07}, which has been shown to have a typical rms uncertainty of $\sim 3\%$ and to be in agreement with the Cepheid scale. As for M31, because of the issue of extinction mentioned above, the $F814W$ TRGB distance reported in EDD is not used here. We take the $F814W$ TRGB magnitude and color in the EDD, and the extinction value at the field of the observation in EDD from the map of \citet{mon09}, then derive its $F814W$ TRGB distance using the same calibration of \citet{riz07} as done in EDD. The resulting distance modulus is then applied to all M31 fields. 
The ACS $F814W$ observations from the PHAT survey cannot be used to derive an $F814W$ TRGB distance for our study because there is no observation in the corresponding fields with $F555W$ or $F606$ filters, which are required for the \citet{riz07} calibration.

\begin{figure}[h]
 \includegraphics[width=\columnwidth,]{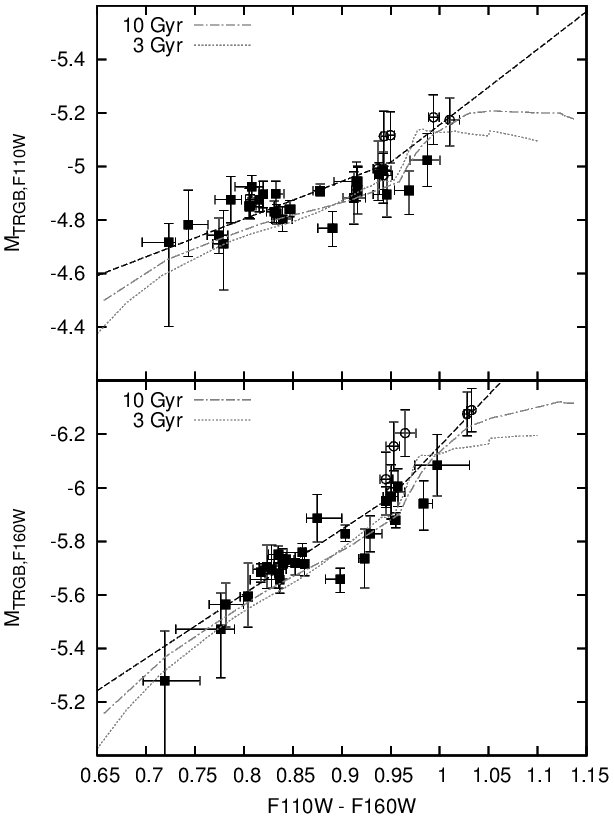} 
\caption{Absolute IR TRGB magnitudes vs. $F110W - F160W$ color. Black dashed lines show the best-fit relations of the color dependencies of TRGB magnitudes. Gray dash-dotted and dotted lines are from isochrone model of \citet{bre12} for ages of 10~Gyr and 3~Gyr with metallicity $0.0001 < Z < 0.06$. The blue end corresponds to the lowest metallicity and the color increases when metallicity increases. Symbols are the same as those in Figure~\ref{fig:814110}. Errors in absolute IR TRGB magnitudes are quadratic sums of errors from TRGB measurements and distance moduli. All data are corrected for extinction. The color dependency in each band is approximated by two linear relations for $F110W - F160W \leq 0.95$ and $F110W - F160W > 0.95$ separately. At $F110W - F160W \leq 0.95$, the color dependency is roughly twice as large in $F160W$ than in $F110W$.}
\label{fig:trgbfit}
\end{figure}

Figure~\ref{fig:trgbfit} plots the absolute IR TRGB magnitudes versus $F110W - F160W$ colors. The errors of magnitudes are quadratic sums of errors from measurements of IR TRGB magnitudes and distance inputs. 
Similar to Figure~\ref{fig:814110}, IR TRGB magnitudes have a nearly linear dependency on color at $F110W - F160W \lesssim 0.95$, but increases rather rapidly at $F110W - F160W \gtrsim 0.95$. Also shown in Figure~\ref{fig:trgbfit} are theoretical predictions of TRGB magnitudes of two different stellar ages with metallicity $0.0001 < Z < 0.06$, from the isochrones of \citet{bre12}. Models predict the upturn starts at $F110W - F160W \sim 0.95$, and the loci are qualitatively consistent with our data. 

Guided by both data and models, we approximate the correlation between IR TRGB magnitude and color in each band by two straight lines with different slopes, where the color demarcation is set to be $F110W - F160W = 0.95$. We fit the relations in two bands simultaneously, with the requirement that the intercepts at $F110W - F160W = 0.95$ of these two bands should differ by 0.95 and the slopes should differ by 1. This requirement is to ensure that at any given color, the two relations give consistent TRGB magnitudes. 

The resulting color dependencies of TRGB magnitudes are:

\tiny
\begin{equation}
 M_{TRGB}(F110W) = \\
  \left\{\begin{array}{r}
         -5.02 - 1.41 \times [(F110W - F160W) - 0.95], \\
                F110W - F160W \leq 0.95 \\
         -5.02 - 2.81 \times [(F110W - F160W) - 0.95], \\
                F110W - F160W > 0.95 .
  \end{array}\right.
\end{equation}

\begin{equation}
\label{eqn:160}
 M_{TRGB}(F160W) = \\
 \left\{ \begin{array}{r}
         -5.97 - 2.41 \times [(F110W - F160W) - 0.95], \\
                               F110W - F160W \leq 0.95 \\
         -5.97 - 3.81 \times [(F110W - F160W) - 0.95], \\
                                 F110W - F160W > 0.95. \\
  \end{array}.\right.
\end{equation}
\normalsize
with an uncertainty on the zero-point of 0.02~mag.

After folding the 0.02~mag uncertainty from the $F814W$ TRGB distances of \citet{riz07} in the quadrature error term, the zero-point uncertainty of IR TRGB distance moduli are 0.03 mag. 

We compute the scatter around each relation separately. The rms uncertainty of IR TRGB distance moduli are

\begin{equation}
\begin{array}{ll}
\sigma_{F110W,blue} = 0.05 \mbox{ mag}, & \sigma_{F110W,red} = 0.12 \mbox{ mag} \\
\sigma_{F160W,blue} = 0.07 \mbox{ mag}, & \sigma_{F160W,red} = 0.09 \mbox{ mag}.
\end{array}
\end{equation}

In practice, to obtain the distance toward a given galaxy, an observation in the halo of the galaxy would be preferred because halo stars usually have low metallicities. By using the lower-metallicity part of the calibration, both the correction of the color term and the rms uncertainty are reduced. We note that, in some elliptical galaxies, the stellar metallicity remains high at large radii \citep{har07a}. Although the uncertainty is higher, the high-metallicity part of the calibration may still be necessary for those galaxies. The other factor that would potentially raise the uncertainty of the TRGB determination is a wide metallicity distribution among stars included in the analysis. This situation could arise with more distant galaxies with strong metallicity gradients, where the WFC3 field of view may cover a large portion of the galaxy therefore a wide metallicity range. In such cases, spatial filtering can effectively isolate stars with a narrower metallicity range. 
However, there would be still cases where a small region in a galaxy contains stars with a wide metallicity range \citep[see][for examples]{har07a,har07b}, where spatial filtering may not work well to isolate stars with similar metallicities. In those rare occasions, analyzing stars from a narrow color slice on the CMD can effectively mitigate the problem of mixing of metallicities \citep[see][for examples drawn from $F814W$ TRGB determinations]{riz07}.

Using the same galaxies in $HST$ Program 11719, \citet{dal12b} approximated the color dependency in $F160W$ as a linear relation with a slope of -2.576. At $0.7 < F110W - F160W < 0.95$, the \citet{dal12b} relation and that of Equation~\ref{eqn:160} agree within $\sim 0.07$ mag. A larger deviation occurs at the red end, where our relation has a steeper slope suggested by the wider color baseline of our calibration sample. At $F110W-F160W=1.05$, the \citet{dal12b} calibration gives a $F160W$ magnitude $\sim 0.15$ mag fainter than Equation~\ref{eqn:160}.

\section{Comparison with $F814W$ TRGB Distances}
\label{sec:compare}

To verify the ability of IR TRGB as a distance indicator, we searched the HST archive for galaxies (1) which are observed in both $F110W$ and $F160W$ by WFC3, (2) whose IR TRGB are above the photometric limit, (3) having enough stars around the magnitude of TRGB for a robust TRGB measurement, and (4) whose $F814W$ TRGB distance is also available. We found 3 galaxies  in addition to our calibration sample fulfill all above requirements. We derive the distance moduli of these galaxies using the color-calibrated absolute TRGB magnitudes. Table~\ref{tbl:compare} lists these galaxies and their TRGB measurements. We also include IC~342, which is observed in both $F606W$ and $F814W$ ACS as well as $F110W$ and $F160W$ by WFC3 in our new $HST$ Program 12877. For IC~342, we use the $F814W$ TRGB distance derived from our new observation instead of the value from EDD. A detailed distance measurement of IC~342 will be presented in Section~\ref{sec:IC342}. 

\begin{table*}
\label{tbl:compare}
\begin{threeparttable}
\tiny
 \caption{$F814W$, $F110W$ and $F160W$ TRGB Distances of the Comparison Sample}
\tabcolsep=0.11cm
\begin{tabular}[t]{llrccccccccccc}
 \hline
\hline
PGC    & Target & T & HST Program & $m_{TRGB}^{F814W}$\tnote{a} & Color\tnote{a} & $m_{TRGB}^{F110W}$ & Color        & $m_{TRGB}^{F160W}$ & Color        & $E(B-V)$\tnote{b} & $m-M$\tnote{a} & $m-M$   & $m-M$   \\
Number &        &  & (Optical,IR) &                    &              &                    &              &                    &           &          & $F814W$        & $F110W$ & $F160W$ \\
\hline
13826 &  IC~342  & 6 & 12877,12877 & 24.697$\pm$0.029 & 2.012$^{+0.015}_{-0.016}$ & 23.183$\pm$0.037 & 1.146$^{+0.003}_{-0.003}$ & 21.979$\pm$0.057 & 1.166$^{+0.003}_{-0.009}$ & 0.541 & 27.76$^{+0.11}_{-0.11}$ & 27.60$^{+0.13}_{-0.13}$ & 27.59$^{+0.14}_{-0.14}$ \\
28655 &  M82    & 8 & 10776\tnote{c},11360 & 24.050$\pm$0.020 & 2.520$^{+0.010}_{-0.020}$ & 22.929$\pm$0.078 & 0.971$^{+0.011}_{-0.001}$ & 21.957$\pm$0.031 & 0.989$^{+0.002}_{-0.001}$ & 0.138 & 27.79$^{+0.02}_{-0.02}$ & 27.75$^{+0.10}_{-0.10}$ & 27.79$^{+0.11}_{-0.11}$ \\
39225 & NGC4214 & 10 & 6569\tnote{d},11360  & 23.330$\pm$0.060 & 1.600$^{+0.010}_{-0.020}$ & 22.356$\pm$0.017 & 0.955$^{+0.005}_{-0.004}$ & 21.343$\pm$0.034 & 0.970$^{+0.004}_{-0.005}$ & 0.019 & 27.30$^{+0.07}_{-0.06}$ & 27.34$^{+0.07}_{-0.07}$ & 27.33$^{+0.11}_{-0.11}$ \\
48082 & M83     & 5 & 10523\tnote{e},11360 & 24.520$\pm$0.030 & 1.420$^{+0.010}_{-0.030}$ & 23.495$\pm$0.101 & 0.998$^{+0.047}_{-0.010}$ & 22.365$\pm$0.062 & 1.009$^{+0.017}_{-0.020}$ & 0.059 & 28.45$^{+0.04}_{-0.03}$ & 28.52$^{+0.21}_{-0.18}$ & 28.43$^{+0.16}_{-0.17}$ \\
\hline
\end{tabular}
\begin{tablenotes}
 \item[a] From EDD, except IC~342. For IC~342, values are from measurements of this work. See Section~5. Values are not corrected for extinction. 
 \item[b] From \citet{sch11}, except IC~342. For IC~342, see Section~5.
 \item[c] $F555W$ and $F814W$ with ACS.
 \item[d] $F555W$ and $F814W$ with WFPC2.
 \item[e] $F606W$ and $F814W$ with ACS.
\end{tablenotes}
\end{threeparttable}
\end{table*}

\begin{figure}
 \includegraphics[width=\columnwidth,]{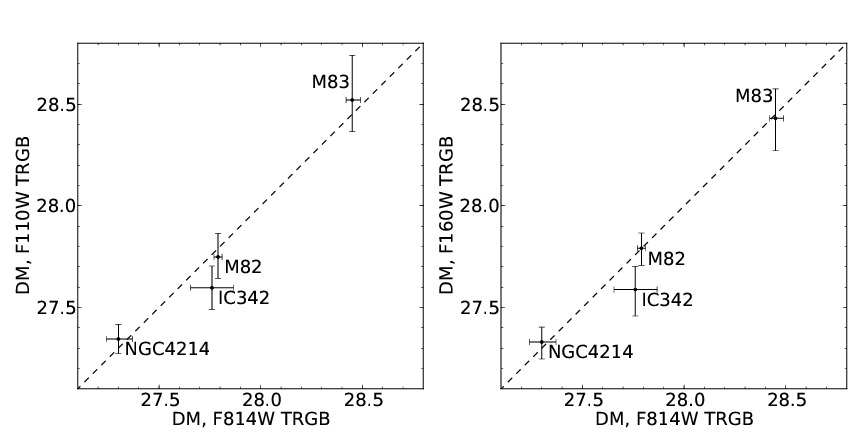}
\caption{Comparison between distance moduli derived from $F814W$ TRGB and IR TRGB. 
Error bars are quadratic sums of rms uncertainties of the calibration and TRGB measurement uncertainties. IR TRGB distances are consistent with $F814W$ TRGB distances.}
\label{fig:compare}
\end{figure}

We compare our IR TRGB distances with $F814W$ TRGB distances in Figure~\ref{fig:compare}. The three galaxies have distances from the two methods consistent with each other to within 1$\sigma$. The biggest offset happens to be IC~342, with the $F110W$ TRGB distance modulus 0.16 mag smaller than the $F814W$ TRGB distance modulus, corresponding to $\sim 8\%$ in distance. Taking uncertainties from both the $F814W$ and the IR TRGB distances into account, this deviation corresponds to $\sim$ 1.0 $\sigma$.

Knowing that the $F814W$ and IR TRGB give consistent distance measurements, which set of single-orbit, 2-filter exposure sequences ($F606W$ and $F814W$ for ACS or $F110W$ and $F160W$ for WFC3) yields a better sensitivity for measuring TRGB magnitudes? Applying selection criteria described in Section~\ref{sec:photometry}, the limiting $F110W$ and $F160W$ magnitudes are $\sim 25.8$ (see Figure~\ref{fig:CMD} for an example) and $\sim 25.0$, respectively. From experience, under the same selection criteria, a single-orbit ACS $F606W$ and $F814W$ exposure sequence yields limiting magnitudes of $\sim 27.6$ in $F606W$ and $\sim 26.8$ in $F814W$ \citep{tul13}.
The magnitude differences between limiting magnitudes in $F814W$ and IR filters are roughly the median $F814W - \mbox{IR}$ color of the TRGB of the calibration sample (see Figure~\ref{fig:814110}). When there is no dust obscuration the two exposure sequences have comparable sensitivity for measuring TRGB magnitudes. The ACS exposure sequence has the advantage on lower metallicity objects because those have bluer TRGB colors, while the WFC3 exposure sequence is more sensitive for higher metallicity objects because of their TRGB is brighter in IR. On the other hand, in case of significant dust extinction, $E(B-V) \gtrsim 0.6$, even for the galaxy with the bluest TRGB color (lowest metallicity) of the calibration sample, the advantage shifts to WFC3 because the magnitude differences between limiting magnitudes of photometry and extincted TRGB magnitude is larger in IR than in optical.

\section{Distances toward the Maffei/IC~342 Group}

\begin{figure}
\includegraphics[width=\columnwidth,]{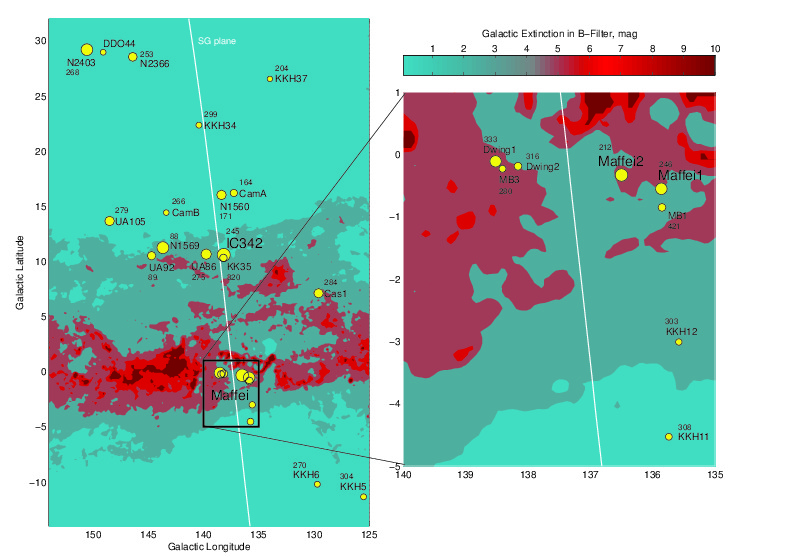}
\caption{Locations of probable members of the Maffei/IC~342 group, on top of the $B$-band Galactic extinction map of \citet{sch98}. Numbers near names of the galaxies indicate their radial velocities (km s$^{-1}$) in the Local Group rest frame. The Maffei/IC~342 group is located at low Galactic latitudes and suffers from severe Galactic extinction, especially Maffei~1 and Maffei~2.}
\label{fig:dust}
\end{figure}

The Maffei/IC~342 group is one of the nearest groups whose integrated luminosity and number of galaxies are comparable to the Local Group. Although this group is important to the dynamics of the local Universe, distances to the dominant group members have been poorly known \citep{kar03,fin03,fin07}. Figure~\ref{fig:dust} shows locations of members of the Maffei/IC~342 group, on top of the Galactic extinction map of \citet{sch98}. Some of the group members are located deep in the Zone of Avoidance and highly obscured by dust in the Galactic plane. With the absolute IR TRGB magnitude calibrated, now the TRGB methodology can be applied to these highly obscured galaxies to derive accurate distances. In this section, we derive distances of the three principal members in the Maffei/IC~342 group. 

Given the small internal uncertainty of the TRGB methodology, extinction dominates the uncertainty in distances of these highly obscured objects. If $E(B-V)$ is under-estimated by 0.1 mag, the true apparent magnitude of the TRGB in $F110W$ is then under-estimated by 0.10 mag. 
Meanwhile, the absolute TRGB magnitude is over-estimated by 0.06 mag due to its color dependence, or 0.10 mag for high metallicity galaxies. In total, the 0.1 mag error on $E(B-V)$ could result in $\sim$0.2 mag error in distance modulus. The error in $F160W$ is similar because, although the apparent magnitude is less affected, the absolute magnitude is more sensitive to color.

\begin{figure}
\includegraphics[width=\columnwidth,]{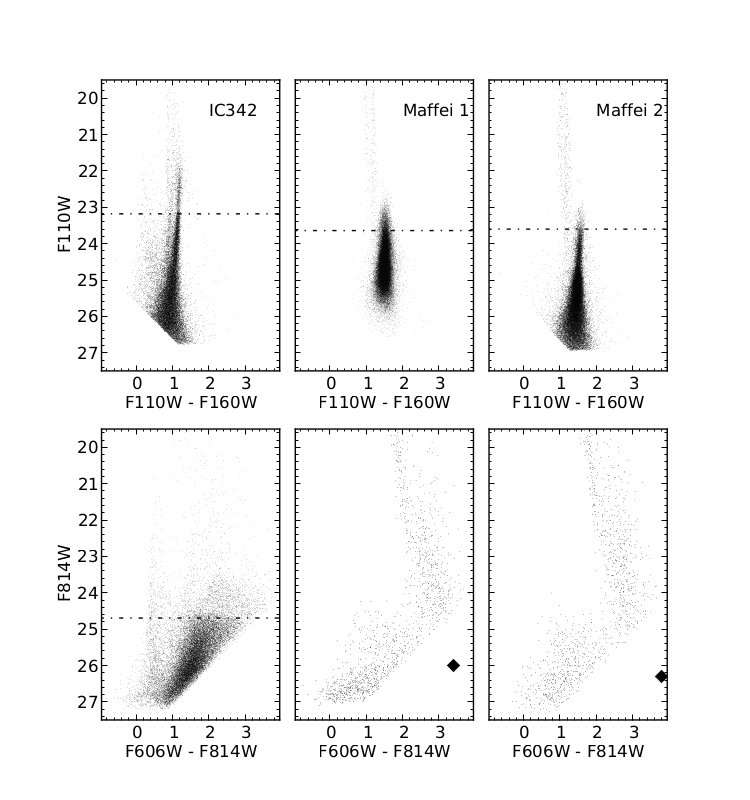}
\caption{IR (upper panels) and optical (low panels) CMDs of IC~342, Maffei~1 and Maffei~2. Data are not corrected for extinction. The TRGB magnitudes are indicated by the horizontal dashed-dotted lines. The optical TRGB for Maffei~1 and Maffei~2 cannot be determined due to severe extinction. The fill diamonds on Maffei~1 and Maffei~2 optical CMDs are expected magnitudes and colors of the optical TRGB, obtained from the isochrone models of \citet{bre12} for an age of 10~Gyr with metallicities estimated from the IR TRGB color, and the input distance moduli are from the IR TRGB distance.}
\label{fig:p12877}
\end{figure}

At low Galactic latitudes, commonly used maps of extinction derived by \citet{sch98} and \citet{sch11} cannot be applied because they are not well calibrated in extreme circumstances. Therefore we derive the extinction of each galaxy individually from the color of specific features of the CMD in order to obtain good TRGB distances. Figure~\ref{fig:p12877} shows $F814W$ v.s. $F606W-F814W$ and $F110W$ v.s. $F110W-F160W$ CMDs of the galaxies observed in Program 12877. In each case, the IR TRGB is identified with small error. We summarize our measurements in Table~\ref{tbl:distance} and explain the procedures used to measure the extinction in this section. Errors in distance moduli are the quadratic sums of errors from zero-point uncertainties, random uncertanties, TRGB measurements, and the dominant reddening component.

\begin{table*}
\label{tbl:distance}
\begin{threeparttable}
 \caption{IR TRGB Distances toward Members of the Maffei-IC~342 Group}
\small
\begin{tabular}{lrcccccccccc}
\hline
\hline
Name & T & \multicolumn{4}{c}{$F110W$} & \multicolumn{4}{c}{$F160W$}     &  $E(B-V)$\\
      \cline{3-6}   \cline{7-10}                   
     & &  $m_{TRGB}$\tnote{a}       & color\tnote{a} & $m-M$   & D (Mpc)    &   $m_{TRGB}$\tnote{a}       & color\tnote{a}  & $m-M$         & D (Mpc) & \\
\hline
IC~342   & 6 & 23.183$\pm$0.037 & 1.146$^{+0.003}_{-0.003}$ & 27.60$^{+0.13}_{-0.13}$ & 3.31$^{+0.20}_{-0.20}$ & 21.979$\pm$0.057 & 1.166$^{+0.003}_{-0.009}$ & 27.59$^{+0.14}_{-0.14}$ & 3.30$^{+0.23}_{-0.22}$ & 0.541$^{+0.066}_{-0.066}$ \\ 
Maffei~1 & -3 & 23.642$\pm$0.074 & 1.499$^{+0.007}_{-0.017}$ & 27.68$^{+0.20}_{-0.17}$ & 3.43$^{+0.32}_{-0.23}$ & 22.097$\pm$0.070 & 1.520$^{+0.015}_{-0.012}$ & 27.71$^{+0.18}_{-0.14}$ & 3.48$^{+0.29}_{-0.21}$ & 1.169$^{+0.046}_{-0.049}$\\
Maffei~2 & 4 & 23.602$\pm$0.037 & 1.554$^{+0.022}_{-0.012}$ & 27.73$^{+0.19}_{-0.19}$ & 3.52$^{+0.32}_{-0.30}$ & 22.021$\pm$0.046 & 1.560$^{+0.016}_{-0.020}$ & 27.69$^{+0.19}_{-0.19}$ & 3.45$^{+0.32}_{-0.29}$ & 1.165$^{+0.080}_{-0.080}$\\
\hline
 
\end{tabular}
\begin{tablenotes}
 \item[a] Numbers are not corrected for extinction. 
\end{tablenotes}
\end{threeparttable}
\end{table*}

\subsection{IC~342}
\label{sec:IC342}

The spiral galaxy IC~342 suffers the least extinction among the three dominant galaxies in the Maffei/IC342 group. It was discovered early in the 1890's \citep{dre95} due to its bright visual magnitude \citep[$V = 8.3$,][]{but99}.

\begin{figure}
 \includegraphics[width=\columnwidth,]{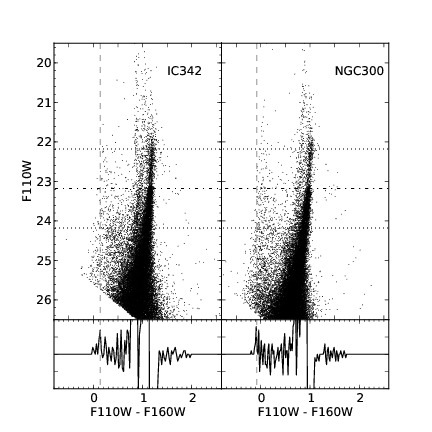}
 \caption{An example of our procedure for determining the color of the blue edge of main sequences. The top panels show CMDs of IC~342 and NGC300. The magnitude of NGC300 is shifted so that the TRGB magnitudes of the two galaxies match. TRGB magnitudes are indicated by the dashed-dotted line. We bin star counts within $\pm$1 mag of TRGB magnitude (dotted lines) in color, then pass the binned color functions through a Sobel kernel of [-2, 0, 2]. The color functions after convolution with the Sobel kernel are shown in the lower panels. The color where the counts change the most, when considering colors bluer than $F110W - F160W \sim 0.5$ , is seen as the maximum and determines the edge of the main sequence (vertical dashed lines).}
 \label{fig:IC342}
\end{figure}

For IC~342, where the TRGB is detected at both optical and IR bands, we derive its reddening by comparing the color of the main sequence to those of other spiral galaxies with small reddening. We adopt the color of the blue edge of main sequence for comparison. For each galaxy, we bin the star counts for stars within $\pm 1$ mag of the $F110W$ TRGB in color. The binned color functions are then passed through a Sobel kernel of [-2, 0, 2]. The output gives a maximum at the color where the count changes the most and is taken as the blue edge of the main sequence.  As examples, Figure~\ref{fig:IC342} shows the CMD and binned color function convolved with the Sobel kernel of IC~342 and NGC~300. The magnitude of the NGC~300 CMD is shifted until the TRGB magnitudes of two galaxies meet. 
To minimize errors, galaxies used to compare with IC~342 should meet the following requirements. First, the galaxy should have a prominent main sequence population, otherwise the color function would be seriously affected by statistical uncertainty. Second, the galaxy should have small reddening so the color uncertainty is minimal. Third, the galaxy should be also observed in $F606W$ and $F814W$ filters so that we could apply this method in optical bands for a consistency check. 
Three galaxies in the calibration sample meet these requirements. Their colors of main sequences' blue edge and $E(B-V)$ are listed in Table~\ref{tab:MScolor}. For each galaxy, the bin size is chosen to be the average photometric error at the magnitude of the TRGB and is considered as the uncertainty of the color. We take the error-weighted average of the de-reddened zero-age main sequence colors of the three galaxies as the intrinsic color of the IC~342 main sequence's blue edge. 

\begin{table*}
\centering
\label{tab:MScolor}
\begin{threeparttable}
\caption{Colors of Blue Edge of Main Sequences and $E(B-V)$ of IC~342}
 \begin{tabular}[t]{lccccc}
\hline
\hline
Target   & \multicolumn{2}{c}{Main Sequence Color} & \multicolumn{2}{c}{De-reddened Main Sequence Color} & $E(B-V)$ \\
         &  $F110W - F160W$ &  $F606W - F814W$ &  $F110W - F160W$ &  $F606W - F814W$ &  \\
\hline
NGC~300   & -0.084$\pm$0.017 & -0.230$\pm$0.028     & -0.089$\pm$0.017 & -0.242$\pm$0.028 & 0.011\tnote{a}  \\
UGC~8508  & -0.094$\pm$0.028 & -0.240$\pm$0.040     & -0.100$\pm$0.028 & -0.254$\pm$0.040 & 0.013\tnote{a}  \\
IC~2574   & -0.095$\pm$0.033 & -0.240$\pm$0.040     & -0.109$\pm$0.033 & -0.275$\pm$0.040 & 0.032\tnote{a}  \\
\hline
Average  &     ---          &    ---               & -0.094$\pm$0.013 & -0.254$\pm$0.020   & --- \\
\hline
IC~342    & ~0.138$\pm$0.025 & ~0.347$\pm$0.063     & -0.095$\pm$0.013 & -0.254$\pm$0.020 & 0.541$\pm$0.066\tnote{b},~0.547$\pm$0.060\tnote{c}  \\
\hline
 \end{tabular}
\begin{tablenotes}
 \item[a] From \citet{sch11}.
 \item[b] Derived from $F110W - F160W$ main sequence color.
 \item[c] Derived from $F606W - F814W$ main sequence color.   
\end{tablenotes}
\end{threeparttable}
\end{table*}

In this manner we obtain $E(B-V) = 0.541 \pm 0.066$ from $F110W$ main sequences and $E(B-V) = 0.547 \pm 0.060$ from $F814W$ main sequences. Although the IR and the optical fields are $\sim 5\arcmin$ apart, the two measurements are consistent with each other. They are also in agreement with the value from \citet{sch11}, $E(B-V) = 0.494 \pm 0.054$, and also close to the value derive from Balmer decrement of \ion{H}{2} regions in IC342, $E(B-V)=0.621 \pm 0.051$ \citep{fin07}.

Applying $E(B-V) = 0.541 \pm 0.066$, the distance modulus from the $F110W$ TRGB is 27.60$^{+0.13}_{-0.13}$ mag, which translates into a distance of 3.31$^{+0.20}_{-0.20}$ Mpc. We note that, as mentioned in Section~\ref{sec:compare}, the distance modulus from the $F814W$ TRGB (27.76$^{+0.11}_{-0.11}$ mag) is 0.16~mag larger, and the corresponding distance (3.56$^{+0.18}_{-0.18}$~Mpc) is $\sim 8\%$ larger. The average distance modulus from $F110W$ and $F814W$ TRGB is 27.69$^{+0.08}_{-0.08}$ mag (3.45$^{+0.13}_{-0.13}$ Mpc) . This distance modulus is in agreement with that from Cepheid period-luminosity (PL) relation, 27.58$\pm$0.18 mag \citep{sah02}. 

\begin{figure}
\includegraphics[width=\columnwidth,]{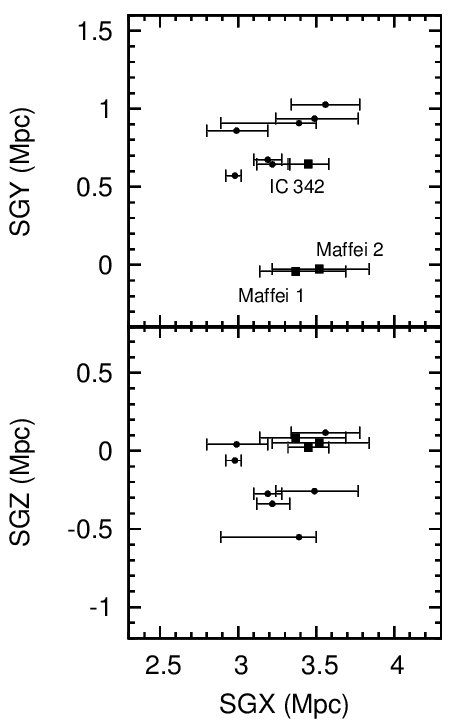}
\caption{Members of the Maffei/IC~342 group with TRGB distances plotted from two orthogonal directions in supergalactic coordinates. IC~342, Maffei~1, and Maffei~2 are represented by filled squares, while other members with known distance are represented by small circles. Distances of IC~342, Maffei~1 and Maffei~2 are from this work, while distances for other galaxies are derived with $F814W$ TRGB, acquired from EDD. The Galactic equator is coincident with SGY=0. Previously measured distances in the group are in less obscured regions in the vicinity of IC~342. Since distances are predominantly aligned with the SGX axis, uncertainties in distances project mainly to this axis.}
\label{fig:sg}
\end{figure}

Several galaxies are located in the projected vicinity of IC~342. Figure~\ref{fig:sg} plots members of the IC~342 group, along with IC~342, Maffei~1, and Maffei~2, in supergalactic coordinates. From 7 members of the IC~342 group, \citet{kar03} reported that the mean distance toward the IC~342 group is $3.28 \pm 0.15$~Mpc. From $F814W$ TRGB distances toward these galaxies in the EDD, which are derived from the calibration of \citet{riz07}, the mean distance toward the IC~342 group is $3.25^{+0.17}_{-0.24}$~Mpc. The mean group distance is also in agreement with the distance of IC~342.

\subsection{Maffei~1}
\label{sec:m1}

Maffei~1 is a giant elliptical galaxy located close to the Galactic plane ($b = -0.^{\circ}55$). The severe extinction makes it appear dim and small at visual wavelength ($V = 11.14$), but if the extinction were removed, Maffei~1 would be among the brightest galaxies in the sky \citep{fin03}). The infrared-bright AGB stars in Maffei~1 are found to be similar to those in other nearby elliptical galaxies such as NGC~5128 and M32, as well as the bulge of M31\citep{dav02}, which suggests that Maffei~1 has an old, metal-rich population.

In Figure~\ref{fig:p12877} we see that Maffei~1 lacks a populated upper main sequence, hence the approach applied on IC~342 for estimating the reddening is not applicable here. Instead, we estimate the reddening by comparing the $F110W - F160W$ TRGB color of Maffei~1 to the bulge of M31, which we assume has a similar stellar content but with less reddening.

In Section~\ref{sec:cali} we measure the TRGB colors of Field~1 in bricks B01 and B05 in the PHAT survey. These two fields are $\sim 1.5$ and $\sim 5.5$ kpc away from the center of M31, respectively. Our WFC3/IR is $\sim$3\arcmin~ away from the center of Maffei~1, which corresponds to $\sim$3~kpc, assuming the distance toward Maffei~1 of 3.5~Mpc. We therefore use the TRGB colors of B01 and B05 as templates for the TRGB color of Maffei~1. 

\begin{figure}
 \includegraphics[width=\columnwidth]{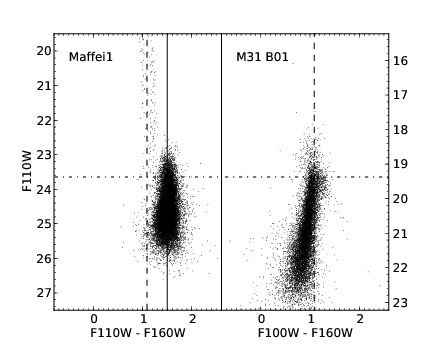}
 \caption{CMDs of Maffei~1 and M31 bulge (B01). The magnitude of M31 B01 is shifted so that the TRGB magnitudes of two galaxies match. TRGB magnitudes are indicated by the dashed-dotted line. The color of Maffei~1 TRGB is indicated by the vertical solid line. Vertical dashed lines show the color of the TRGB of M31 B01.}
 \label{fig:maffei1}
\end{figure}

Figure~\ref{fig:maffei1} shows CMD of Maffei~1 and B01 of M31. The CMD of B01 is shifted in magnitude so that the TRGB magnitude matches the TRGB magnitude of Maffei~1. Although the two CMDs have similar shapes, the TRGB color of Maffei~1 is $\sim 0.4$ mag redder. For B01 and B05 of M31, the de-reddened $F110-F160W$ TRGB colors are $1.021^{+0.010}_{-0.002}$ and $1.006^{+0.006}_{-0.005}$, respectively. 
We assume the TRGB color of Maffei~1 equals the average TRGB color of B01 and B05 (1.013). For the error budget, we take the high end of B01 (1.031) and the low end of B05 (1.001) as the $1 \sigma$ color uncertainty of Maffei~1, in order to account for possibly imperfect match between the age/metallicity of stellar populations in M31 bulge and Maffei~1.

With this assumption, we obtain the reddening toward Maffei~1 $E(F110W-F160W)_{Maffei~1} = 0.498^{+0.020}_{-0.021}$, therefore $E(B-V)_{Maffei~1} = 1.169^{+0.046}_{-0.049}$. 
We check whether the extinction varies across the WFC3 field by comparing the colors of the Galactic main sequence, the strips extending to bright magnitudes in Maffei~1 CMDs, in different sub-regions. We do not find perceptible color differences across the WFC field.

We note that a few other studies had attempted to measure the extinction towards Maffei~1. By comparing the nuclear spectrum with early-type galaxies, \citet{spi71} adopted a value of $A_V = 5.2 \pm 0.2$ mag. \citet{nan73} studied the distribution of M stars towards Maffei~1 and a neighboring comparison region and concluded that the extinction $A_V$ should be between 4 and 5 mag. A later study by \citet{but83} measured the extinction in two ways. First, by comparing the $B-V$ color via aperture photometry of Maffei~1 and other unobscured giant ellipticals, the $A_V$ was estimated to be $5.3 \pm 0.4$ mag. Second, the total column density of gas in the direction of Maffei~1, measured from both H~I and CO yielded $A_V = 4.9 \pm 0.4$. More recent measurements gave slight lower $A_V$ values of $4.5 \pm 0.8$ and $4.67 \pm 0.19$, estimated from the $H - K$ color of AGB stars \citep{dav02} and the relation between the Mg$_2$ index and $V-I$ color \citep{fin03,fin07}, respectively.

Our value of $E(B-V)_{Maffei~1} = 1.169$ translates into $A_V = 3.74$, which is about 1 mag lower than values in the literature. Assuming $4.5 \lesssim A_V \lesssim 5.0$ as reported, the reddening is then $ 0.61 \lesssim E(F110W-F160W) \lesssim 0.67$, which implies that the true TRGB color of Maffei~1 is $ 0.83 \lesssim (F110W-F160W)_{TRGB} \lesssim 0.89$. For Maffei~1, a metal-rich giant elliptical galaxy, such a blue TRGB color would not be expected. This fact suggests that previous studies could over-estimate the extinction towards Maffei~1.

\begin{figure}
 \includegraphics[width=\columnwidth]{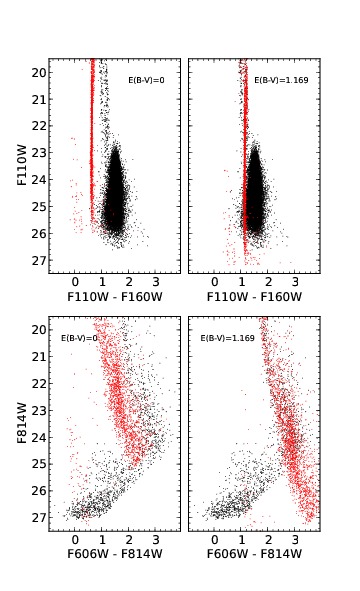}
 \caption{Comparison between observed and simulated Galactic main sequence stars in the Maffei~1 field. Real data are plotted in black, without correcting for extinction. Simulated Galactic stars generated by TRILEGAL code are plotted in red. The $E(B-V)$ in each panel is the $E(B-V)$ value applied to the simulated stars. With our estimate $E(B-V) = 1.169$, simulated and observed Galactic main sequences agree with each other in both IR and optical wavelength.}
 \label{fig:m1ext}
\end{figure}

To further verify our extinction estimate, we compare the Galactic main-sequence appearing in Maffei~1 CMD with simulated Galactic stars generated by the TRILEGAL code\footnote{http://stev.oapd.inaf.it/cgi-bin/trilegal} \citep{gir12}. The simulated population contains several Galactic components: thin disk, thick disk, halo, and bulge. Each of them has its own geometry, age-metallicity relation, and star formation rate. We choose default settings of TRILEGAL, which are calibrated by \citet{gir05} and \citet{van09}. The initial mass function is also set to be the default of TRILEGAL, the Chabrier log-normal function \citep{cha01}. The simulated area is chosen to match the field of view of either ACS or WFC3.
In Figure~\ref{fig:m1ext}, we take $E(B-V)=1.169$, apply corresponding extinction in each band to the simulated un-extincted Galactic main-sequence stars generated by TRILEGAL, then plot them on top of Maffei~1 CMDs. We find that the observed Galactic main-sequence agrees with simulated main-sequence with our estimated $E(B-V)$. Therefore we adopt our reddening estimate rather than values in the literature.

Given the assumed extinction, the distance modulus and distance towards Maffei~1 from the $F110W$ TRGB are $27.68^{+0.20}_{-0.17}$ and $3.43^{+0.32}_{-0.23}$ Mpc, respectively. This measurement can be compared with previous estimates of distance toward Maffei~1 that have a wide dispersion. \citet{lup93} used the $K'$-band surface brightness fluctuation technique, derived a distance of $4.15 \pm 0.5$ Mpc. \citet{dav01} compared the $K$-band magnitude of the brightest AGB stars in Maffei~1 and in the bulge of M31, conclude a distance of $4.4^{+0.6}_{-0.5}$ Mpc. Through the fundamental plane in $I$-band, \citet{fin03} estimated a closer distance of $3.01\pm0.30$ Mpc, and later revised it with a new distance zero point, $2.85\pm0.36$ Mpc \citep{fin07}. Our IR TRGB distance locates Maffei~1 in the middle of previous estimates.

\subsection{Maffei~2}

The barred spiral galaxy Maffei~2 is located at low Galactic latitudes ($b = -0.^{\circ}33$). The Galactic extinction towards Maffei~2 was estimated to be even more severe than Maffei~1 \citep{spi73,fin07}. Although its apparent $V$ magnitude, $V = 12.41$, is by about 1 mag dimmer than that of Maffei~1, after correcting for extinction, Maffei~2 is potentially as equally dominant as Maffei~1 in the IC342/Maffei group \citep{fin07}.

The extinction correction for Maffei~2 is particularly ambiguous. Maffei~1 is expected to be metal rich so it is reasonable to assume Maffei~1 has among the reddest of intrinsic TRGB colors. IC~342 has a prominent main-sequence population so the color-stable main sequence feature of the CMD is used to derive the extinction. In the case of Maffei~2, the main sequence is not well populated in the halo field that was observed, so the color of the main sequence cannot be reliably derived by the same method as it was for IC~342.

Lacking features that characterize reddening attributed to stars within Maffei~2, we turn to the foreground stars in the CMD. The Galactic main-sequence appears in the Maffei~2 field, as it does with Maffei~1. We therefore compare the narrow Galactic main sequence in IR with TRILEGAL to estimate the extinction.

\begin{figure}
 \includegraphics[width=\columnwidth]{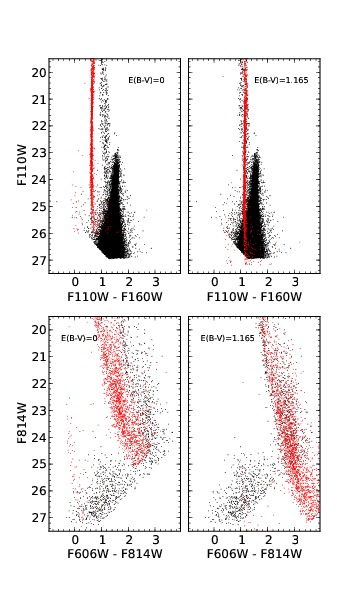}
 \caption{Comparison between observed and simulated Galactic main sequence stars in the Maffei~2 field. Real data are plotted in black, without correcting for extinction. Simulated Galactic stars generated by TRILEGAL code are plotted in red. The $E(B-V)$ in each panel is the $E(B-V)$ value applied to the simulated stars. With our estimate $E(B-V) = 1.165$, simulated and observed Galactic main sequences agree with each other in both IR and optical wavelength.}
 \label{fig:m2ext}
\end{figure}

First, we fit a straight line to the magnitude-color relation of simulated Galactic main sequence over a selected magnitude range, $20 \leq F110W \leq 22.5$. We choose this magnitude range because (1) the observed Galactic main sequence merges with stars of Maffei~2 at $F110W \simeq 24$, which corresponds to an un-extincted $F110W$ magnitude of $\sim 22.5$ with the expected extinction and (2) the color dispersion of simulated Galactic main sequence becomes larger at $F110W \gtrsim 20$ therefore a straight line may not be representative. Second, we take $E(B-V)$ as a free parameter, apply the extinction to the magnitude-color relation of the simulated Galactic main sequence, then fit the extincted simulated main sequence to the observed Galactic main sequence stars with $21.5 \leq F110W \leq 24$. We find the best-fitted $E(B-V)=1.165$ and show the result in Figure~\ref{fig:m2ext}. We estimate the uncertainty on $E(B-V)$ by varying the magnitude range of observed stars included in the fitting process. 
Fixing the faint magnitude limit of $F110W = 24$, we change the bright magnitude limit from 20.5 to 22.5. This results in a range of reddening, $1.09 \lesssim E(B-V) \lesssim 1.24$. Therefore we assign an uncertainty of 0.08~mag to our best-fitted $E(B-V)$.

We find the extinction towards Maffei~2 in the literature is markedly higher than the value we derived here. For example, the extinction derive by \citet{fin07}, using Balmer decrement of \ion{H}{2} regions in Maffei~2, corresponds to $E(B-V) \simeq 1.75$. If we applying this $E(B-V)$ value to Maffei~2, the de-reddened IR TRGB color will be $\sim 0.8$. For Maffei~2, a Sbc ($T=4$) galaxy, such a blue TRGB color is not expected. In the calibration sample, galaxies with TRGB color $\sim 0.8$ are mostly irregulars with $T=10$. On the other hand, normal spirals have IR TRGB color $\gtrsim 0.9$ (see Table~\ref{tbl:galaxydata}). With $E(B-V)=1.165$ the de-reddened TRGB color of Maffei~2 becomes $1.054^{+0.040}_{-0.036}$, which is more consistent with what would be expected from its morphological type. With $E(B-V)$ derived, the $F110W$ distance modulus and distance toward Maffei 2 is $27.73^{+0.19}_{-0.19}$ and $3.52^{+0.32}_{-0.30}$ Mpc. 
Here the error is mainly from the uncertain extinction, which results in $\sim 0.14$ mag uncertainty from apparent and absolute magnitudes combined. 

Previously, only few studies have estimated the distance to Maffei~2. Through the $B$-band Tully-Fisher relation, \citet{kar03} estimated a distance of 2.8~Mpc. More recent studies, also through the Tully-Fisher relation, but in $I$ and $K_s$-band, \citet{fin07} and \citet{kar10} derived distances of $3.34\pm0.56$ and 3.1~Mpc, respectively. Our value is close to what have been derive from the Tully-Fisher relation.

\begin{figure}
 \includegraphics[width=\columnwidth]{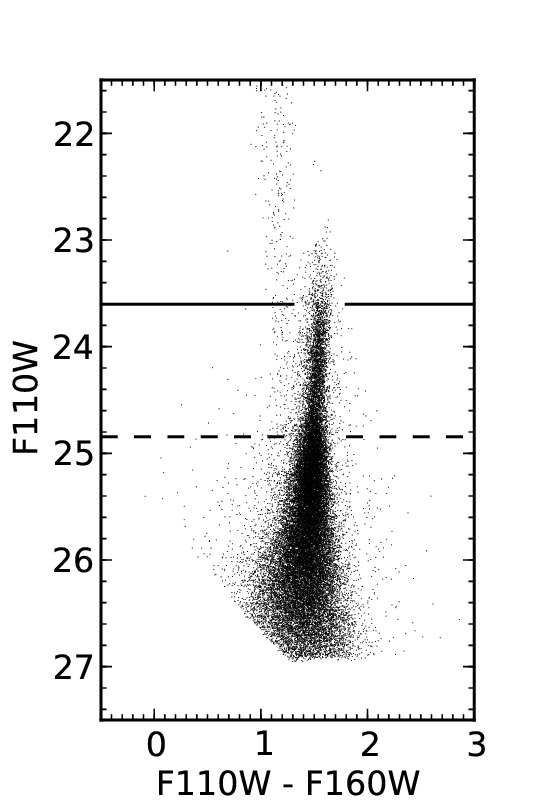}
 \caption{Maffei~2 CMD. The TRGB adopted in this paper is labeled by the break solid lines at $F110W = 23.602$, while a second discontinuity is found at $F110W = 24.845$, labeled by the break dashed lines.}
 \label{fig:m2cmd}
\end{figure}

We would like to point out that we identify a second discontinuity in the luminosity function of Maffei~2. Figure~\ref{fig:m2cmd} shows the CMD of Maffei~2, with its TRGB magnitude indicated by the solid line at $F110W = 23.602$. It is recognizable that the density of stars rises at $F110W \sim 25$ and we detect a discontinuity at $F110W = 24.845$, indicated by the dashed lines in Figure~\ref{fig:m2cmd}. One possibility is that we mistake the tip of the AGB population as the TRGB and the true TRGB is at the fainter discontinuity. If this is the case, Maffei~2 would be located at $> 6$ Mpc, which is less likely based on previous knowledge of the distance towards Maffei~2. However, if we take the extinction $A_V = 5.6$ from \citet{fin07}, despite that this value give an unlikely blue TRGB color, combined with TRGB at $F110W = 24.845$, the resulting distance is 3.6~Mpc, a coincidentally plausible value.

For both Maffei~1 and Maffei~2, the big uncertainty of TRGB distances, and essentially all other distance measures, are largely contributed from the uncertain extinction estimates. In the near future, data from surveys covering low Galactic latitudes, such as PanSTARRS \citep{kai02} or Gaia \citep{per01}, will be able to provide better extinction estimates \citep{han14}.

\section{Summary}

In this paper, we calibrate the color dependency of the TRGB magnitudes in \textit{HST} WFC3 $F110W$ and $F160W$ filters. IR TRGB provides an alternative to the commonly-used $I$-band with the benefits that the TRGB is brighter in the IR and dust extinction is reduced. In each band, we approximate the color dependencies of the TRGB magnitudes by two linear relations for low and high metallicity regimes respectively. In spite of a stronger color dependency, the TRGB magnitudes at IR wavelengths still provide good distance measures. The distance from the IR TRGB method yields a $\sim$5\% relative uncertainty (extinction aside) and is given a zero point that agrees with $F814W$ TRGB distances. At the high metallicity regime, possibly due to the line-blanketing effect, the color dependency is stronger and the rms uncertainties of the calibration are larger. We therefore suggest using the low metallicity regime for distance determinations, by obtaining observations toward the halos of target galaxies.

We demonstrate that the IR TRGB method has an advantage over the $F814W$ TRGB method when the Galactic dust extinction is severe. Using the IR TRGB method, we derive distances toward three principal galaxies in the Maffei-IC~342 complex: IC~342, Maffei~1 and Maffei~2. These galaxies suffer from severe Galactic dust extinction, especially, Maffei~1 and Maffei~2, whose $F814W$ TRGB magnitudes are not detectable in our observations. With the IR TRGB method, new distance estimates from the $F110W$ TRGB are 3.45$^{+0.13}_{-0.13}$~Mpc for IC~342 (after averaging the $F814W$ TRGB distance), 3.43$^{+0.32}_{-0.23}$~Mpc for Maffei~1 and 3.52$^{+0.32}_{-0.30}$~Mpc for Maffei~2. The dominant source of uncertainty is the uncertain Galactic extinction, especially for Maffei~1 and Maffei~2. 
In the near future, with surveys such as Gaia or PanSTARRS covering low Galactic latitudes, a more accurate reddening map is possible. Combining with IR TRGB method, the accuracy of distances to objects located in the Zone of Avoidance can be improved. The next generation space telescope, $JWST$, will work at only IR wavelengths. With the IR TRGB magnitudes calibrated, once $JWST$ is operational, the power of TRGB method will be enhanced dramatically. 

\acknowledgments

We thank the anonymous referee for thorough and constructive comments.  
I.K. acknowledges support from RFBR grants Nos. 12-02-91338 and 13-02-92690. This work is based on observations made with the NASA/ESA Hubble Space Telescope, obtained from the data archive at the Space Telescope Science Institute. STScI is operated by the Association of Universities for Research in Astronomy, Inc., under NASA contract NAS 5-26555. Support for program GO-12877 was provided by NASA through a grant from the STScI.
This research has made use of the NASA/IPAC Extragalactic Database (NED) which is operated by the Jet Propulsion Laboratory, California Institute of Technology, under contract with the National Aeronautics and Space Administration.

\end{document}